\documentclass[%
 aip,
 amsmath,amssymb,
 reprint,%
]{revtex4-1}

\usepackage{graphicx}
\usepackage{dcolumn}
\usepackage{bm}
\usepackage[utf8]{inputenc}
\usepackage[T1]{fontenc}
\usepackage{mathptmx}
\usepackage{etoolbox}
\usepackage{textcomp,nicefrac}
\usepackage{url}
\usepackage{comment}

\makeatletter
\def\@email#1#2{%
 \endgroup
 \patchcmd{\titleblock@produce}
  {\frontmatter@RRAPformat}
  {\frontmatter@RRAPformat{\produce@RRAP{*#1\href{mailto:#2}{#2}}}\frontmatter@RRAPformat}
  {}{}
}%
\makeatother
\begin{document}


\title{First Demonstration of Underground Muon Imaging at an Archaeological Site in Ancient Jerusalem}
\author{Y. Benhammou}
 \affiliation{School of Physics and Astronomy, Tel-Aviv University}

\author{E. Etzion$^*$}
\email{ereze@tau.ac.il}
 \affiliation{School of Physics and Astronomy, Tel-Aviv University}

\author{Y. Gadot}%
\affiliation{ 
Department of Archaeology and Ancient Near Eastern Cultures, Tel-Aviv University
}%

\author{O. Lipschits}%
\affiliation{ 
Department of Archaeology and Ancient Near Eastern Cultures, Tel-Aviv University
}%
\author{G. Mizrachi}
 \affiliation{School of Physics and Astronomy, Tel-Aviv University}

 \author{Y. Shalev}
\affiliation{%
Israel Antiquities Authority, Israel
}%

\author{Y. Silver}
 \affiliation{School of Physics and Astronomy, Tel-Aviv University}
 \affiliation{%
Rafael Advanced Defense Systems LTD, Israel
}%

\author{A. Weissbein }
\affiliation{%
Rafael Advanced Defense Systems LTD, Israel
}%

\author{I. Zolkin}
 \affiliation{School of Physics and Astronomy, Tel-Aviv University}

\date{\today}

\begin{abstract}
We present a novel underground imaging system that utilizes cosmic-ray muons to explore the subsurface environment at the City of David archaeological site in ancient Jerusalem. The method exploits the fact that muons lose energy as they travel through matter, with attenuation depending on the integrated density along their path. By tracking muon trajectories through a multi-layered, scintillator-based detector, we reconstruct angular flux distributions and infer variations in overburden density. This report details initial findings from measurements conducted at a large rock-hewn installation, commonly known as “Jeremiah’s cistern.” A high-resolution LiDAR scan of the interior was combined with muon flux simulations to map structural anomalies. The system successfully identifies variations in ground opacity, demonstrating the viability of muon tomography for archaeological imaging in complex  environments.
This work represents a significant interdisciplinary effort to deepen our understanding of this historical site.
\end{abstract}

\maketitle

\section{Introduction}
\begin{figure}
    \centering
    \includegraphics[width=0.8\linewidth]{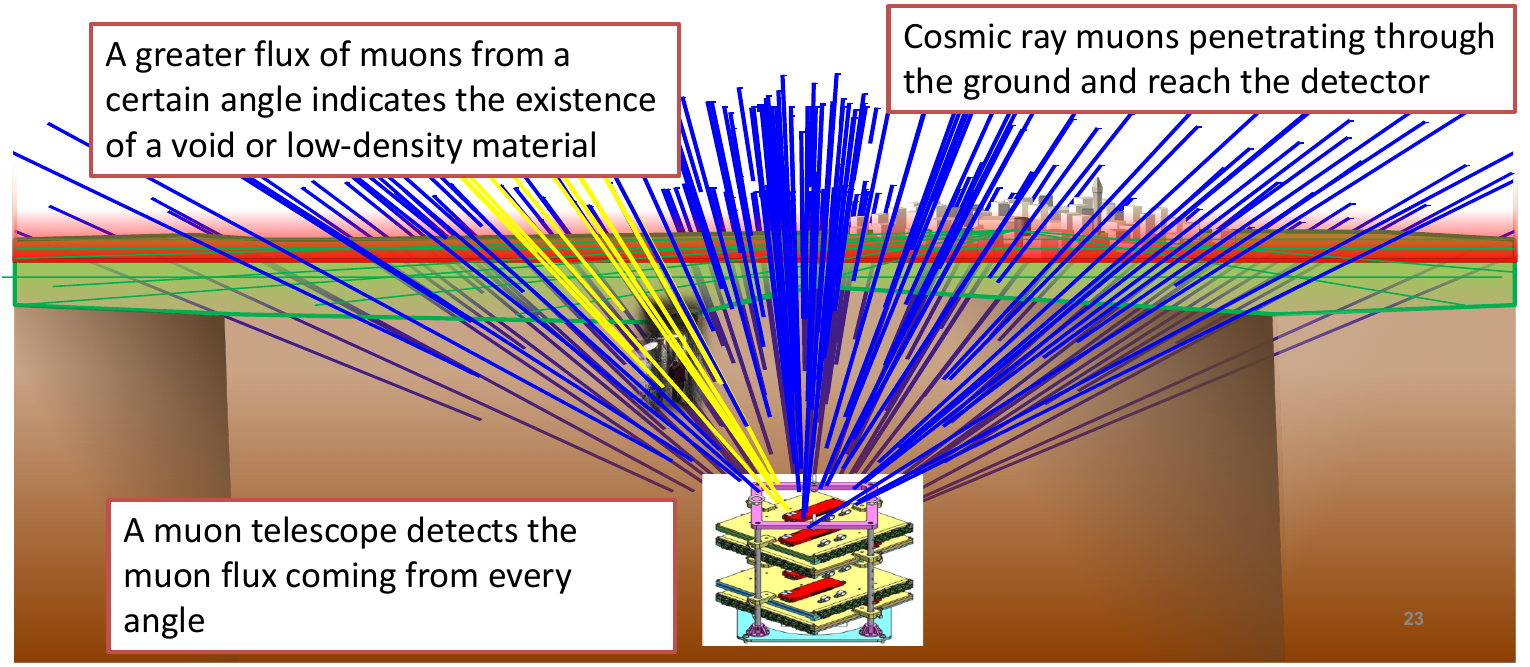}
    \caption{Schematic illustration of cosmic-ray muons penetrating the overburden from various angles. The ground is continuously bombarded by muons at a known rate and angular distribution. As muons lose energy while traversing matter, their flux is attenuated depending on the integrated density along their path. A greater-than-expected flux from a particular direction indicates reduced attenuation, suggesting the presence of a void or low-density region (marked for illustration as yellow lines). The muon telescope detects these angular flux variations to map subsurface structures.}
    \label{fig:muon-penetration}
\end{figure}

Muon tomography is a non-invasive imaging technique that exploits cosmic-ray muons to probe the density distribution within a target material~\cite{Alvarez_1970, Nagamine_2008, Schouten:2018kla}. Muons are highly penetrating particles produced by interactions of cosmic rays with atoms in the upper atmosphere. At ground level, their flux is approximately 1 muon per cm$^2$ per minute, or about 10,000 muons per m$^2$ per minute, with energies ranging from a few hundred MeV to several TeV. Unlike X-rays, muons can traverse tens to hundreds of meters of rock or concrete, making them ideal probes for dense or inaccessible structures. As they pass through matter, muons lose energy via ionization and radiative processes, and their trajectories are altered by scattering and absorption. The level of attenuation depends on the integrated density along their path. By measuring the rate and direction of muons that pass through a target volume using several tracking detectors, a tomographic image of the internal structure can be reconstructed. This is illustrated in Fig.~\ref{fig:muon-penetration}
In our system, muons are detected directly through the ionization they produce while traversing plastic scintillator bars, which emit light in response to deposited energy. The typical duration of data collection ranges from a few days to several weeks, depending on detector size, overburden thickness, and the desired resolution. The muon flux is generally stable but may exhibit mild environmental dependence, particularly on atmospheric pressure and temperature, which affect muon production altitudes and energy spectra.
This technique has recently been applied in geological and archaeological investigations~\cite{TANAKA2003657, Schouten:2018kla, MURAVES:2020lgx}.

The south-eastern ridge (also known as the City of David) in ancient Jerusalem, with stratified deposits dating back to the second and first millennia BCE~\cite{Reich}, provides an excellent test bed for muon tomography. This work describes the first operational test of our custom-developed muon detector deployed at Jeremiah’s cistern (referenced in Jeremiah 38:6), a site adjacent to ongoing excavations. Our objectives include (i) demonstrating the efficacy of muon tomography for archaeological imaging, (ii) performing muon angular rates measurements, (iii) deriving integrated opacity - the integrated ground density traversed by muons (calculated by  $\int{\rho \cdot dl} $ where $\rho$ is the ground density and the integration is performed along the muon path) and (iv) preparing for a second campaign, in which the detector 
will be placed underneath the City of David near the Gihon spring.  

The cistern is a large, rock-hewn installation. A picture of the cistern is shown in Fig.~\ref{fig:Pitt_photo_BW}.
The feature, approximately six meters deep, includes a narrow shaft and a bell-shaped lower chamber, exhibits typical water system characteristics and was likely used for water storage or containment. 
The age of the cistern is still unclear. Nevertheless, its location—at the top of the ridge, adjacent to the "Large Stone Structure" and the "Rock-Cut Moat", two major architectural features interpreted by some scholars as part of a royal construction project, suggests that it may have been part of a prominent building complex. These structures were uncovered in Area G of the City of David excavations and are associated with the First Temple period, which generally refers to the era between the construction of the First Temple in Jerusalem and its destruction (circa 1000–586 BCE)~\cite{Mazar, Mazar2020, Gadot}.

\begin{figure}[h!]
\centerline{\includegraphics[width=0.45\textwidth]{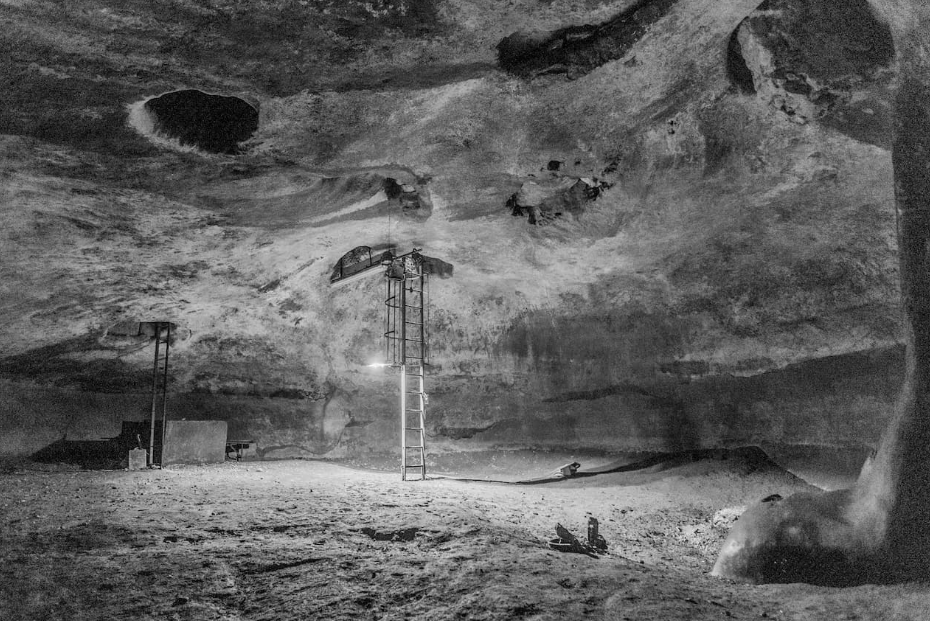}}
\caption{Photograph of the cistern, the bell-shaped water facility underneath the City of David visitor center.}
\label{fig:Pitt_photo_BW}
\end{figure}

\begin{figure}[h!]
\centerline{\includegraphics[width=0.45\textwidth]{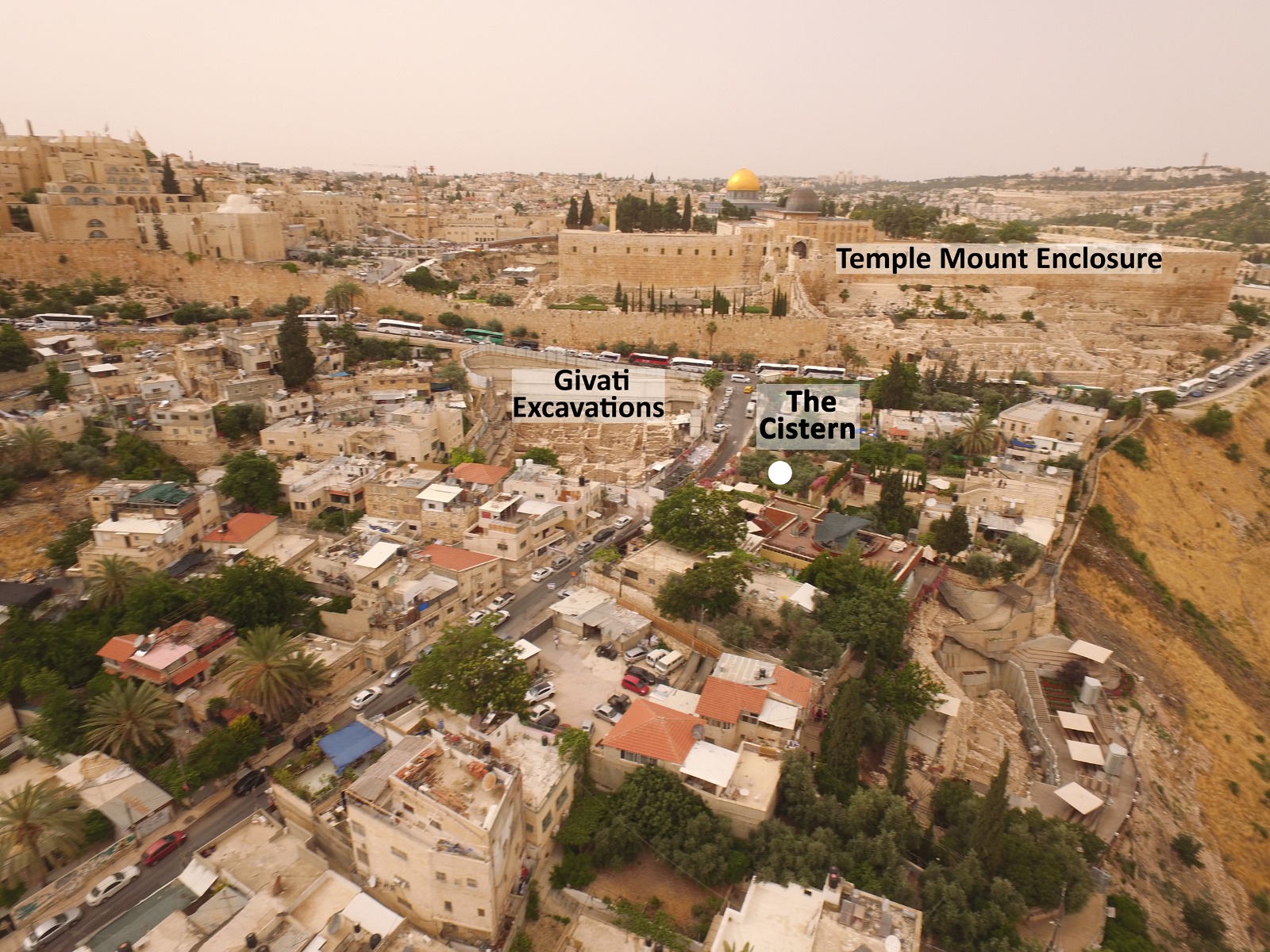}}
\caption{The location of the cistern on the northern part of the City of David (looking north). The location of the cistern is marked. (Photograph by Yair Isboutsky, Courtesy of City of David Archive). }
\label{fig:Pitt_photo_BW}
\end{figure}

\section{Experimental Setup and Methods}
\subsection{Detector Design and Calibration}
The muon detector is a compact, scintillator-based telescope optimized for underground applications. It comprises four detection layers containing 23 extruded plastic scintillator bars, 40~cm long, with a triangular cross-section of \(3.3 \times 1.7\,\text{cm}^2\). 
The triangular cross-section enables sub-pixel position resolution by analyzing the light distribution between adjacent bars, as detailed in Section~\ref{II.C} and illustrated in Fig.~\ref{fig.TriangleBarExplanation}.
The scintillators are fabricated from polystyrene doped with PPO (2,5-Diphenyloxazole) and POPOP (1,4-bis(5-phenyloxazol-2-yl)benzene), which act as primary and secondary fluors, respectively, and are coated with titanium dioxide to enhance light reflection (see Ref.~\cite{Pla-Dalmau:2000puk}). When traversed by a muon, scintillation light is produced and captured by wavelength-shifting (WLS) fibers (Saint-Gobain BCF-91A). 
These fibers absorb the blue scintillation light ($\sim$420 nm) emitted by the plastic scintillator and re-emit it in the green ($\sim$500 nm), a wavelength range with lower attenuation and better matched to the spectral sensitivity of the Sensel microJ 30035 Silicon Photomultipliers (SiPMs)~\cite{Onsemi} mounted on a Printed Circuit Board (PCB). This wavelength shifting improves light collection and signal uniformity over the fiber length. While the re-emission process introduces a small timing delay and may lead to mild spatial inhomogeneities in light yield, these effects are negligible in our application and do not bias the reconstruction performance.
Each fiber extends an additional 50~cm beyond the scintillator to mitigate attenuation losses. A detailed description of the detector is given in Ref.~\cite{Benhammou:2022nry}.

Calibration of the detector involved measuring individual scintillator responses, gain matching for the SiPMs and validation of the detector's sensitivity uniformity.
The detector was initially calibrated in a controlled laboratory environment at Tel Aviv University using cosmic-ray muons. Individual scintillator responses were measured and the gains of the SiPMs were equalized. Based on these tests, the trigger threshold was set to 2600 Analog-to-Digital Converter (ADC) counts, above the electronic noise floor. However, this value was close to the most probable value (MPV) of a muon signal, measured at 3096 ADC counts with a distribution width of approximately 250 counts (this corresponds to a pulse height of roughly 100 mV for a muon passing through a triangular bar near its center). This is shown in Fig. 6 of the  publication describing the detector~\cite{Benhammou:2022nry}. This configuration, while effective at suppressing noise, also rejected a substantial fraction of valid muon signals, resulting in a detection efficiency of approximately 30\%. For subsequent data acquisition, the threshold was lowered and a stricter event selection criterion was applied: valid events were required to register hits in all four scintillator layers. This improved detection efficiency to close to 90\%, while maintaining negligible noise contamination.

\subsection{Data Acquisition System}
The detector employs the CAEN DT5550W data acquisition system based on the WeeROC Application-Specific Integrated Circuit (ASIC)~\cite{CAEN}. A mezzanine card with four 32-channel Citiroc ASICs, combined with a Xilinx XC7K160T~\cite{xilinx} FPGA (Field-Programmable Gate Array), processes the signals. A 14-bit, 80~MS/s ADC digitizes the signals in real time. The system is powered by 128 bias voltage lines (20--35~V) and housed in a dedicated cooling box to ensure thermal stability.

\begin{figure}[h!]
\centerline{\includegraphics[width=0.35\textwidth]{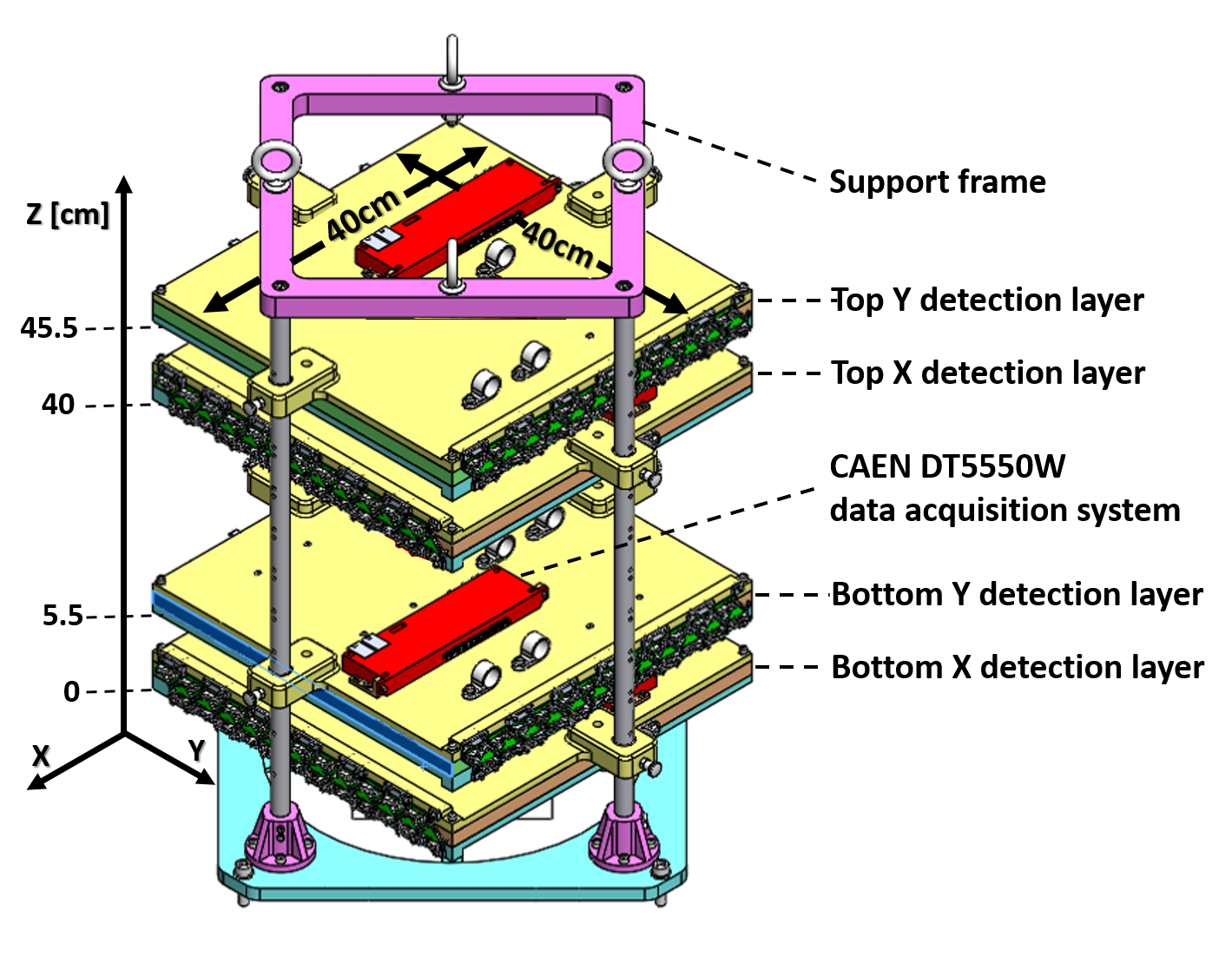}}
\centerline{\includegraphics[width=0.48\textwidth]{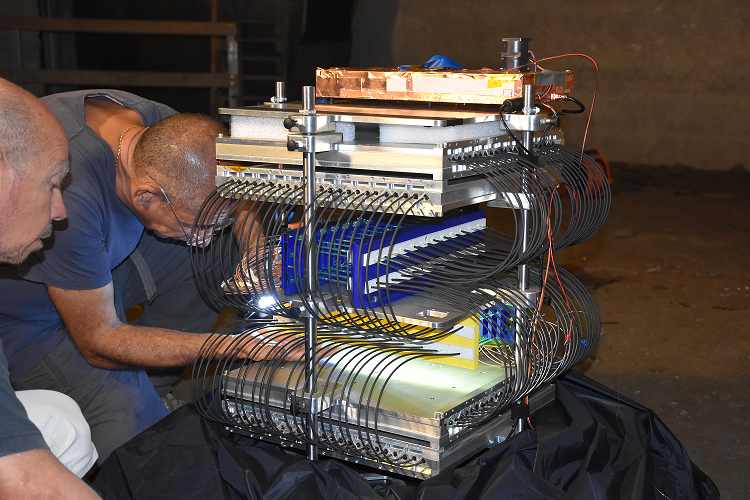}}
\caption{{\bf (Top):} An artist’s representation of the detector, showing the main elements and relevant sizes, the optical fibers were omitted for clarity. {\bf (Bottom):} The muon detector during its commissioning inside the cistern.}
\label{fig.Detector}
\end{figure}

Fig.~\ref{fig.Detector} shows an artist’s representation of the detector with relevant sizes and a photograph of the detector during installation and commissioning in the cistern.

\subsection{Muon Trajectory Determination}
\label{II.C}
Each cosmic-ray muon is expected to leave a trace in all four scintillator layers. The overall muon trajectory is reconstructed by correlating the hit positions across these layers. However, accurate determination of the hit location within each layer is nontrivial due to the geometry of the scintillator bars and the strong dependence of the signal amplitude on both the angle of incidence and the transverse impact position.

In particular, the path length of a muon within a single triangular scintillator bar varies significantly even for vertically incident particles. If the muon passes near a vertex, the path length may approach zero, while a hit near the center of the bar yields the maximum path length, equal to the triangle’s height ($\approx$1.7 cm). This geometric effect causes substantial variation in signal amplitude even at a fixed angle, making direct position estimation from single-bar signals unreliable.

\begin{figure}[htb]
    \centering
    \includegraphics[width=0.8\linewidth]{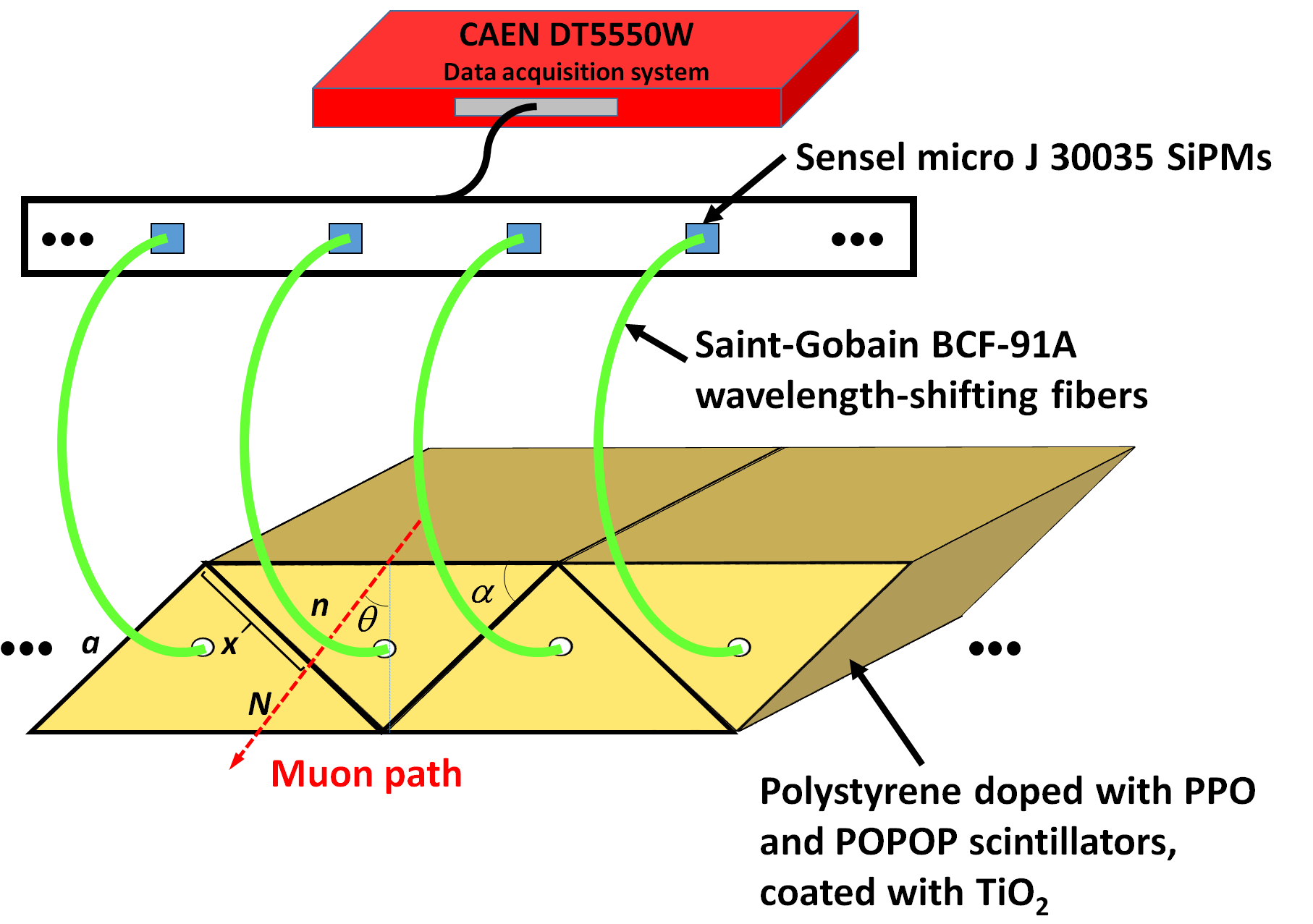}
\caption{Illustration of a muon passing through two adjacent triangular scintillator bars. Due to the triangular geometry, the muon's path length, and thus the signal amplitude, varies strongly with its transverse position, even at normal incidence. By comparing the signal amplitudes \(n\) and \(N\) collected in adjacent bars, the hit position \(X\) can be estimated using a known geometric relation (Eq.~\ref{eq:SubPixelEq}). This method enables sub-pixel spatial resolution, essential for accurate trajectory reconstruction. Also shown schematically are the WLS fibers from the scintillator to the SiPM and the data acquisition.}
    \label{fig.TriangleBarExplanation}
\end{figure}

To overcome this, our reconstruction algorithm exploits the relative light yield in adjacent scintillator bars. 
When a muon crosses the interface between two triangular bars, it traverses a path of length \(X\) in one bar and \(a\) - \(X\) in the other, where \(a\) is the base width of the triangular cross-section and \(X\) is measured from the vertex. Since the amount of scintillation light produced is proportional to the path length, the collected signals \(n\) and \(N\) in the two bars satisfy:

$n \propto X$, $\quad N \propto a - X$

Taking the ratio of the two signals yields:

\begin{equation}
    \frac{n}{N} = \frac{X}{a - X}.
    \label{eq:SubPixelEq}
\end{equation}
Assuming that the signal amplitudes follow Poisson statistics (\(\Delta n = \sqrt{n}\) and \(\Delta N = \sqrt{N}\)), the uncertainty in \(X\) is given by error propagation:
\begin{equation}
    \delta X^2 = \left(\frac{\partial X}{\partial N}\Delta N\right)^2 + \left(\frac{\partial X}{\partial n}\Delta n\right)^2,
\end{equation}
which yields:
\begin{equation}
    \delta X = \frac{a}{N+n}\sqrt{\frac{n \cdot N}{N+n}}.
    \label{eq:SubPixelErrorEq}
\end{equation}
This method, illustrated in Fig.~\ref{fig.TriangleBarExplanation}, enables sub-pixel resolution and is essential for achieving the detector’s overall angular resolution of approximately 20 mrad. 
This resolution corresponds to a spatial resolving power of 1–2 meters at typical overburden distances of 50–100 meters, which matches the expected scale of archaeological features such as cisterns, shafts, and stone alignments.
It also inherently accounts for variation in signal amplitude caused by the muon’s transverse position and angle, thereby improving spatial accuracy and the consistency of track reconstruction.

For overall trajectory reconstruction, the refined hit positions from each layer are correlated. Denote the hit positions in the top and bottom layers as \(x_{\text{top}}\) and \(x_{\text{bottom}}\) (and similarly for \(y\)). The differences \(\Delta x = x_{\text{bottom}} - x_{\text{top}}\) and \(\Delta y = y_{\text{bottom}} - y_{\text{top}}\) are computed over a known vertical separation \(\Delta z\). The muon incident angles are then given by:
\begin{equation}
\tau_x = \tan(\theta_x) = \frac{\Delta x}{\Delta z}, \quad \tau_y = \tan(\theta_y) = \frac{\Delta y}{\Delta z}.
\end{equation}

\subsection{Flux Measurement and Angular Ground Depth Estimation}
Muon trajectories are binned in the angular space \((\tau_x, \tau_y)\) to form the count matrix \(C_{\mu}(\tau_x, \tau_y) \), and Poisson statistics govern the uncertainty in each bin.

The muon rate per angular bin is defined as:
\begin{equation}
R_{\mu}(\tau_x, \tau_y) = \frac{C_{\mu}(\tau_x, \tau_y)}{T_{\mathrm{det}}},
\end{equation}
with \(T_{\mathrm{det}}\) being the exposure time.
It is common to calculate and represent the muonic flux in spherical coordinates (as it depends only on the zenith angle $\theta$ and not the azimuthal angle $\phi$) this is shown in the next sections. In order to correlate the measured flux with the theoretical expectation, we calculated the differential muon flux using: 

\begin{equation}
F_{\mu}(\tau_x, \tau_y) = \frac{R_{\mu}(\tau_x, \tau_y)}{S_{\mathrm{eff}}(\tau_x, \tau_y)\, J},
\end{equation}
where \(S_{\text{eff}}(\tau_x, \tau_y)\) is the effective detection surface, defined as the geometrical area of the detector that contributes to muon detection in direction \((\tau_x, \tau_y)\), weighted by detection efficiency and angular acceptance. It accounts for both the physical size and directional sensitivity of the detector. \(J = \cos^3\theta\) is the Jacobian for the transformation from \((\tau_x,\tau_y)\) to spherical angles \((\theta,\phi)\).
The angular domain ($\tau_x$, $\tau_y$) was divided into uniform square bins, in dimensionless tangent units, in both directions. This bin size was chosen to balance angular resolution with statistical significance per bin. While the binning is uniform in $(\tau_x, \tau_y)$, it corresponds to non-uniform solid angles in spherical coordinates due to the Jacobian transformation. This effect is corrected in the calculation of the differential flux using the factor $J = \cos^3\theta$. No angular-dependent biases were observed beyond expected statistical fluctuations.

\subsection{Estimation of Integrated Opacity}
\label{II.E}
Our objective is to use the measured angular muon rate to estimate the angular ground depth \(D(\tau_x, \tau_y)\). This reflects the effective overburden thickness above the detector.
As noted earlier, the muon flux is reduced with ground depth, because muons lose energy while traversing through the ground until they stop and decay.
The muon flux is related to the integrated density (or opacity) \(\rho L\) (in g\,cm\(^{-2}\)) along the muon path via energy-dependent attenuation:
\begin{equation}
    \frac{dE}{dX} = -a(E) - b(E)\cdot E
\end{equation}

Here, \(a(E)\) represents nearly constant ionization losses (approximately \(2\,\mathrm{MeV\,cm^2/g}\), per the Bethe-Bloch formula) and \(b(E)\) represents radiative losses (including bremsstrahlung, pair production, and photonuclear interactions) that become significant at higher energies~\cite{Groom:2001kq, ParticleDataGroup:2024cfk}. The integrated opacity is obtained by solving:
\begin{equation}
\int_{E_{\mathrm{min}}}^{E_{\mathrm{GL}}} \frac{dE}{a(E) + b(E)\,E} = \rho L,
\label{eq.loss}
\end{equation}
where \(E_{\mathrm{GL}}\) is the muon energy at ground level and \(E_{\mathrm{min}}\) is the threshold energy for muon reaching a specific ground depth. Detailed muon transport simulations yield lookup tables correlating the measured flux with \(\rho L\).

For this, we simulated the expected muonic flux using the parameterization for the vertical intensity shown in ~\cite{Reyna2006ASP} and ~\cite{Bugaev1998}. Specifically, the muon intensity at any zenith angle $\theta$ and muon momentum $p_\mu$ is expressed as

\begin{equation}
I(p_\mu, \theta) = \cos^3\theta\, I_v\bigl(p_\mu\cos\theta\bigr),
\label{eq:combined1}
\end{equation}

where the vertical muon intensity \(I_v(p_\mu)\) is given by

\begin{equation}
I_v(p_{\mu}) = c_1\, p_{\mu}^{-1 {\bigl(c_2 + c_3\log_{10}(p_{\mu}) + c_4\log_{10}^2(p_{\mu}) + c_5\log_{10}^3(p_{\mu})\bigr)}}
\label{eq:combined2}
\end{equation}

Using the best-fit coefficients~\cite{Reyna2006ASP} \(c_1 = 0.00253\), \(c_2 = 0.2455\), \(c_3 = 1.288\), \(c_4 = -0.2555\), and \(c_5 = 0.0209\),  parameterization accurately reproduces the observed muon intensities over a broad momentum range. Equations~\eqref{eq:combined1} and \eqref{eq:combined2} were incorporated into our simulation framework to predict the angular and momentum distribution of muons at the surface, which was subsequently used to estimate the expected muonic flux for every angle and depth. The results of these calculations were combined into a comprehensive lookup table that relates expected muon rates at every angle for every effective depth.

\subsection{Uncertainty and Error Analysis}

Quantifying uncertainties reliably is crucial for resolving ground structure deviations and ensuring the robustness of the reconstruction.
The main sources of uncertainty include:
\begin{itemize}
    \item \textbf{Detector Efficiency:} The overall efficiency (less than 20\%)—due to high trigger thresholds and non-optimal trigger logic—introduces systematic bias.
    \item \textbf{Atmospheric Modeling:} Our simulation used sea-level muon flux models; however, Jerusalem's altitude (approximately 650~m) yields a higher flux and slightly altered angular distribution.
    \item \textbf{Simulations:} Limitations in the muon transport simulations, mainly due to insufficient modeling of the surface and terrain, contribute additional uncertainties. This is described in section~\ref{II.E}.
    \item \textbf{Statistical Fluctuations:} Poisson statistics dictate the uncertainty in each angular bin.
\end{itemize}
The follow-up work aims to optimize trigger thresholds, incorporate altitude corrections, and refine simulation models, which is expected to significantly reduce the main sources of uncertainty.

\section{Results}
After thorough calibration and extensive laboratory testing, the detector was deployed at the cistern in the City of David archaeological site. A high-resolution Light Detection and Ranging (LiDAR) scan was used to map the interior geometry of the cistern, including the vertical shafts. This structural model was then incorporated into our simulation framework to predict the expected muon flux under the assumption of a homogeneous overburden and flat surface.

\subsection{Integrated Muon Rate and Overall Efficiency}
\begin{figure}[h!]
\centerline{\includegraphics[width=0.49\textwidth]{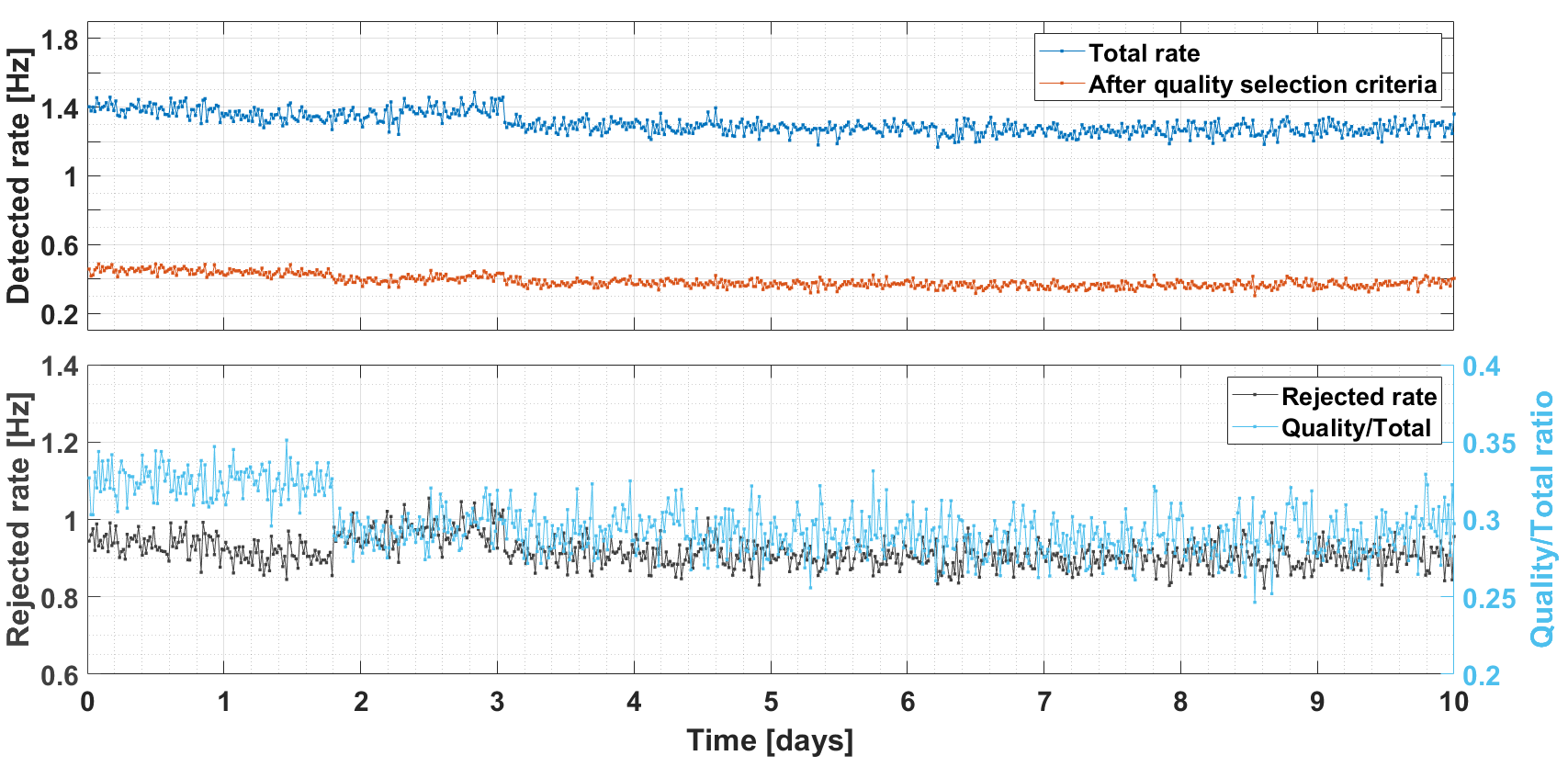}}
\caption{Muon rates during the measurement, recorded continuously over ten days and grouped into 20-minute bins for display.
 {\bf (Top):} Total detected muon rate (blue) and the subset passing strict quality selection criteria (orange), reflecting overall detector performance.
{\bf (Bottom):} Rejected muon rate (black), calculated as the difference between total and quality-selected events, and the corresponding quality/total rate ratio (turquoise, right axis), which reflects the relative tracking efficiency.
The data remain stable throughout the acquisition period, with  mild fluctuations in rate and efficiency. This dual representation highlights both the absolute rejection level and the relative selection behavior over time.}
\label{fig:muon_rate}
\end{figure}

 The detector was deployed and operated at the cistern for a total duration of approximately  ten days (14,709 minutes). The total muon rate \( \bigl( \int_{\tau_x}\int_{\tau_y} R_{\mu}(\tau_x, \tau_y) \bigr) \) and the rate after applying quality selection criteria acquired during this period is shown in Fig.~\ref{fig:muon_rate}. 
We have used only those muon tracks that were analyzed with super resolution (i.e., recorded in two adjacent bars in each of the four detector layers). These constitute our quality selection criteria, ensuring sub-pixel spatial resolution and consistent angular reconstruction.

 The expected muon rate of approximately 2.45 Hz was calculated by combining surface muon flux models from Reyna~\cite{Reyna2006ASP} and  Bugaev~\cite{Bugaev1998} with muon transport simulations based on the energy loss formalism described in Eq.~\ref{eq.loss}. This corresponds to an expected total of approximately 2.12 million muons over the full exposure period. By comparing these measured total rates to the expected value, we estimate the overall detector efficiency to be approximately 50\% for all detected muons, and only 15–20\% for the subset of muons passing the quality selection. The discrepancy between observed and simulated rates is thus partially explained by the stricter selection criteria applied to the data, which were not included in the simulations. Future refinements will integrate these criteria into the simulation chain for more accurate efficiency modeling.

\subsection{LiDAR Mapping and Angular Muon  Distribution}
\begin{figure}[h!]
\centerline{\includegraphics[width=0.45\textwidth]{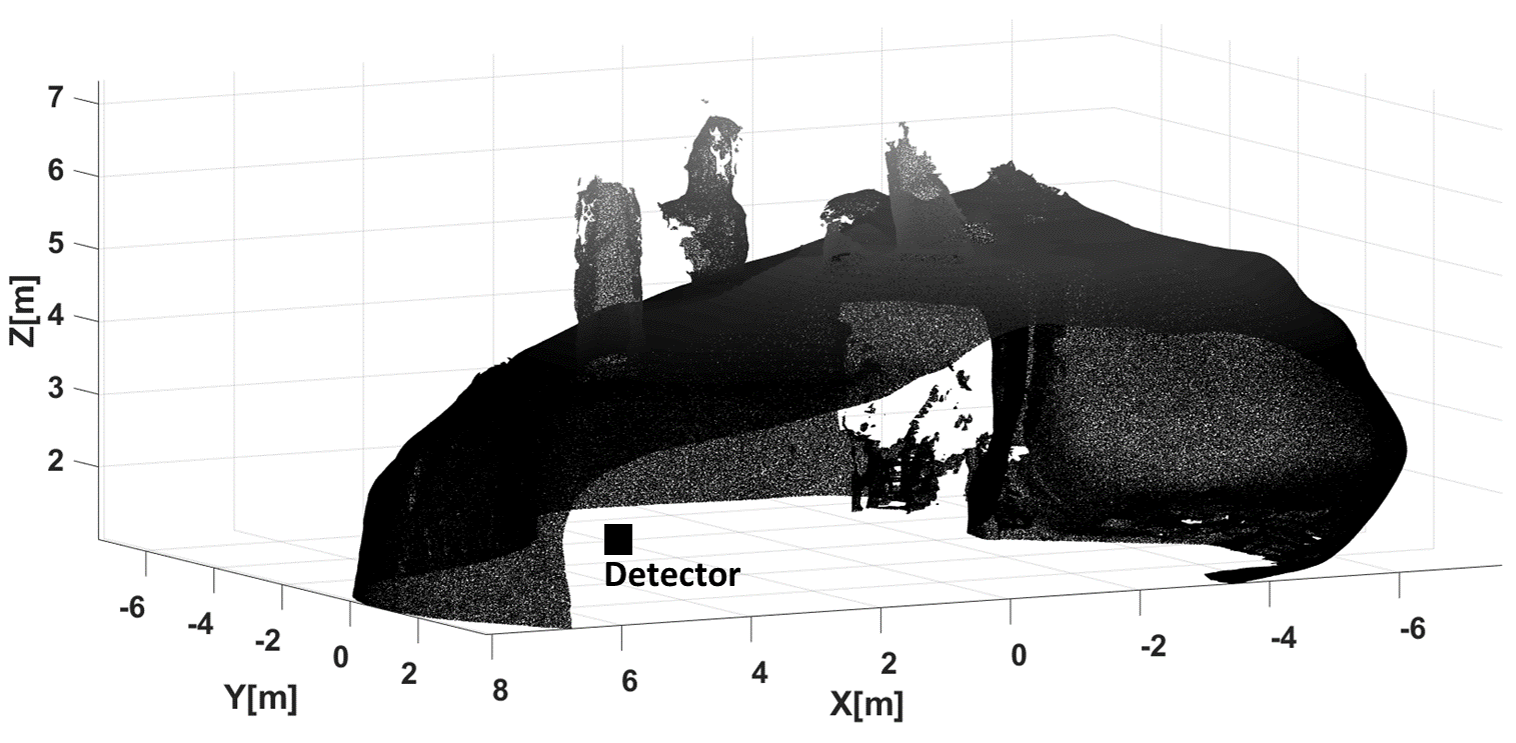}}
\caption{LiDAR image of the interior of the cistern, showing also the ventilation shafts and the detector's location.}
\label{fig:pit}
\end{figure}

\begin{figure}[ht]
\centering
\includegraphics[width=0.45\textwidth]{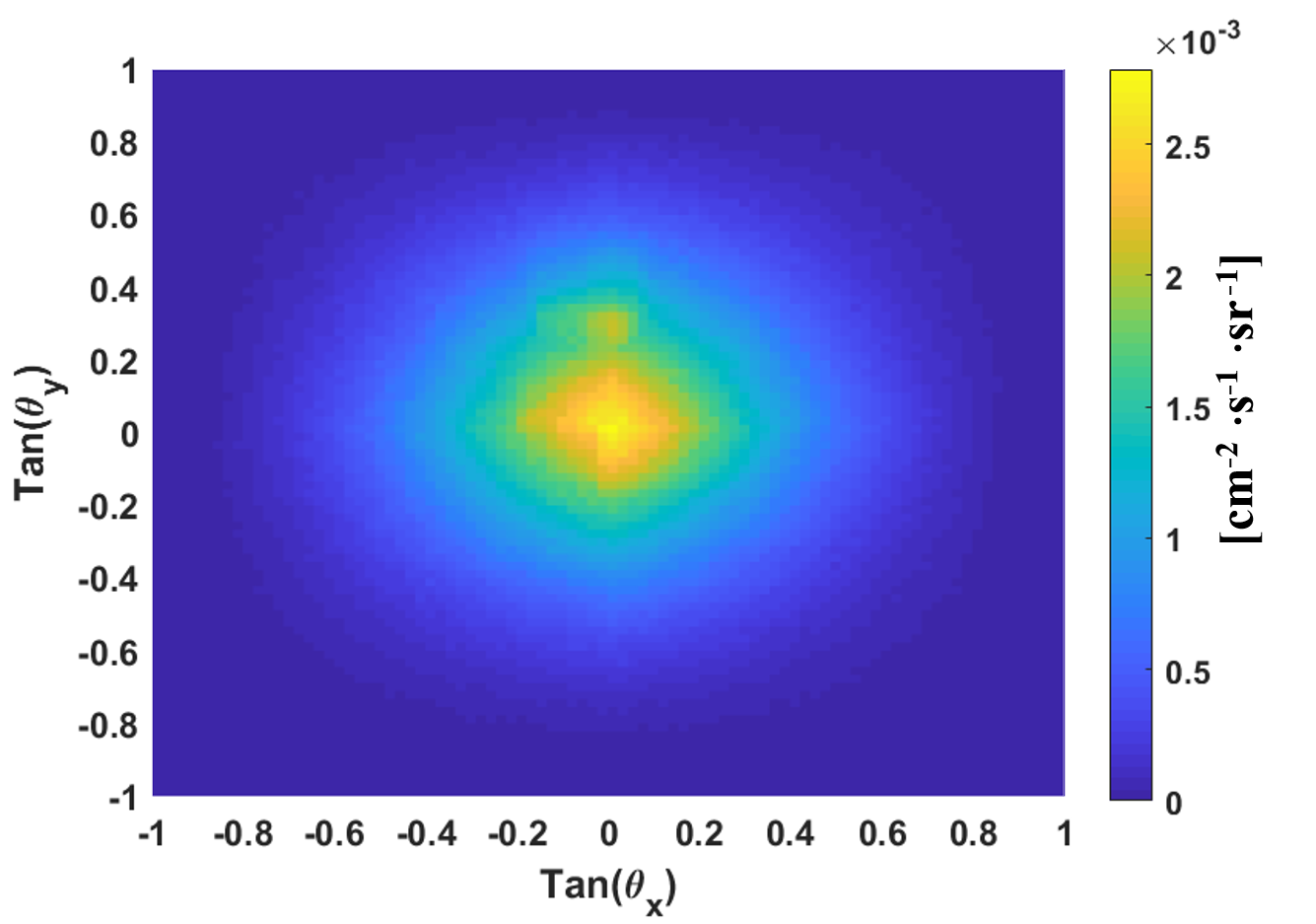}
\includegraphics[width=0.45\textwidth]{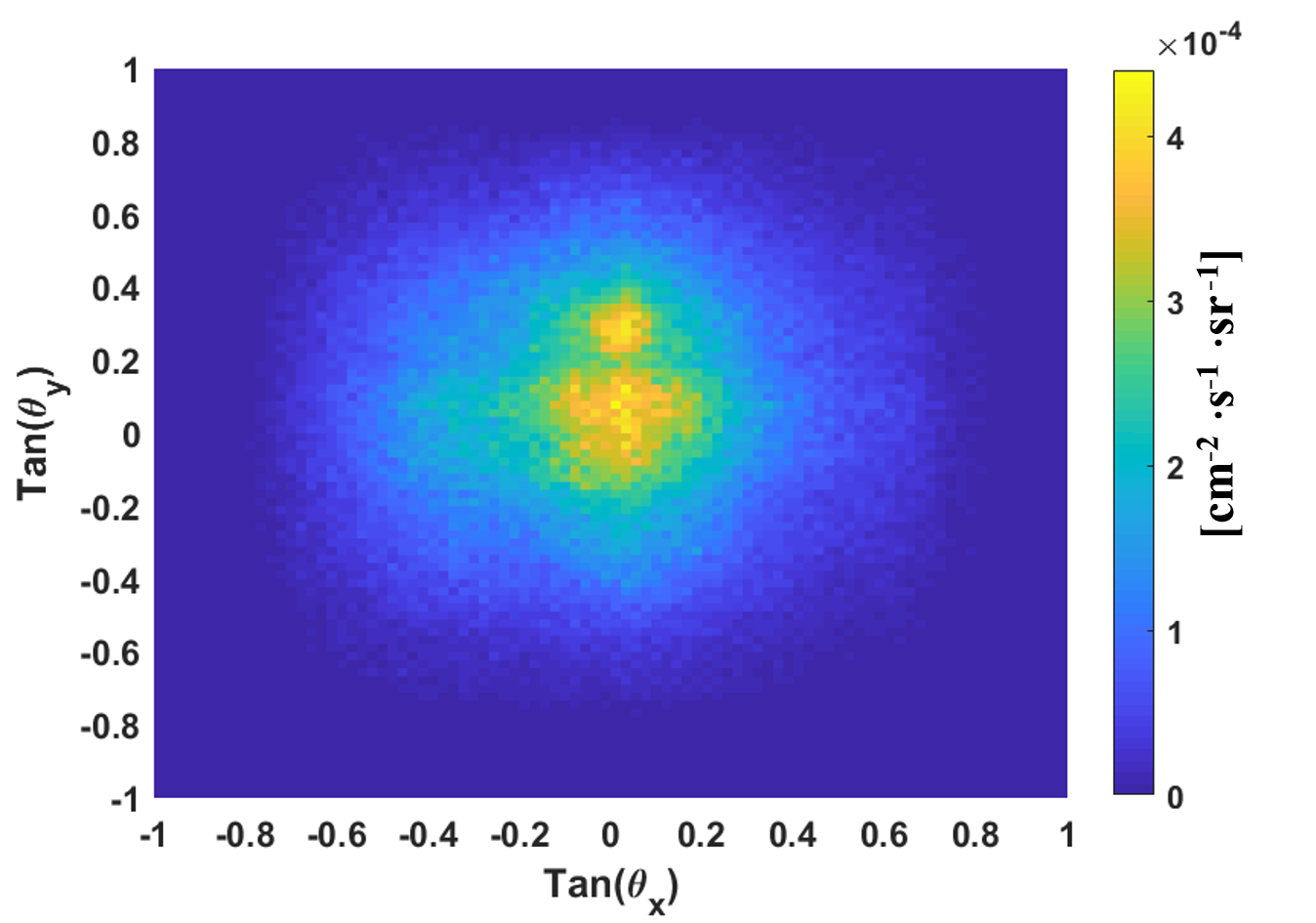}
\caption{\(\tan\theta_x\) versus \(\tan\theta_y\) distributions.  {\bf (Top):}  Muon rates derived from the simulated flux multiplied by the detector's effective surface. {\bf (Bottom):} Measured muon rate data.}
\label{fig:2}
\end{figure}

Fig.~\ref{fig:pit} shows a LiDAR image of the interior of the cistern, clearly revealing the ventilation shafts extending toward the ground surface and the detector's location. Fig.~\ref{fig:2} presents a qualitative comparison between the \(\tan\theta_x\) versus \(\tan\theta_y\) distributions of the muon rates. The top panel shows simulation results for the expected angular muon rate (the simulated flux $\times$ the detector's effective surface), while the bottom displays the measured angular muon rate. Although the absolute numbers differ significantly (discussed in the previous section), both distributions exhibit a maximum near the zenith, with a distinct enhancement corresponding to the known ventilation shaft.

\subsection{Measured Angular Ground Depth}
\begin{figure}[htb]
  \centering
  \includegraphics[width=0.5\textwidth]{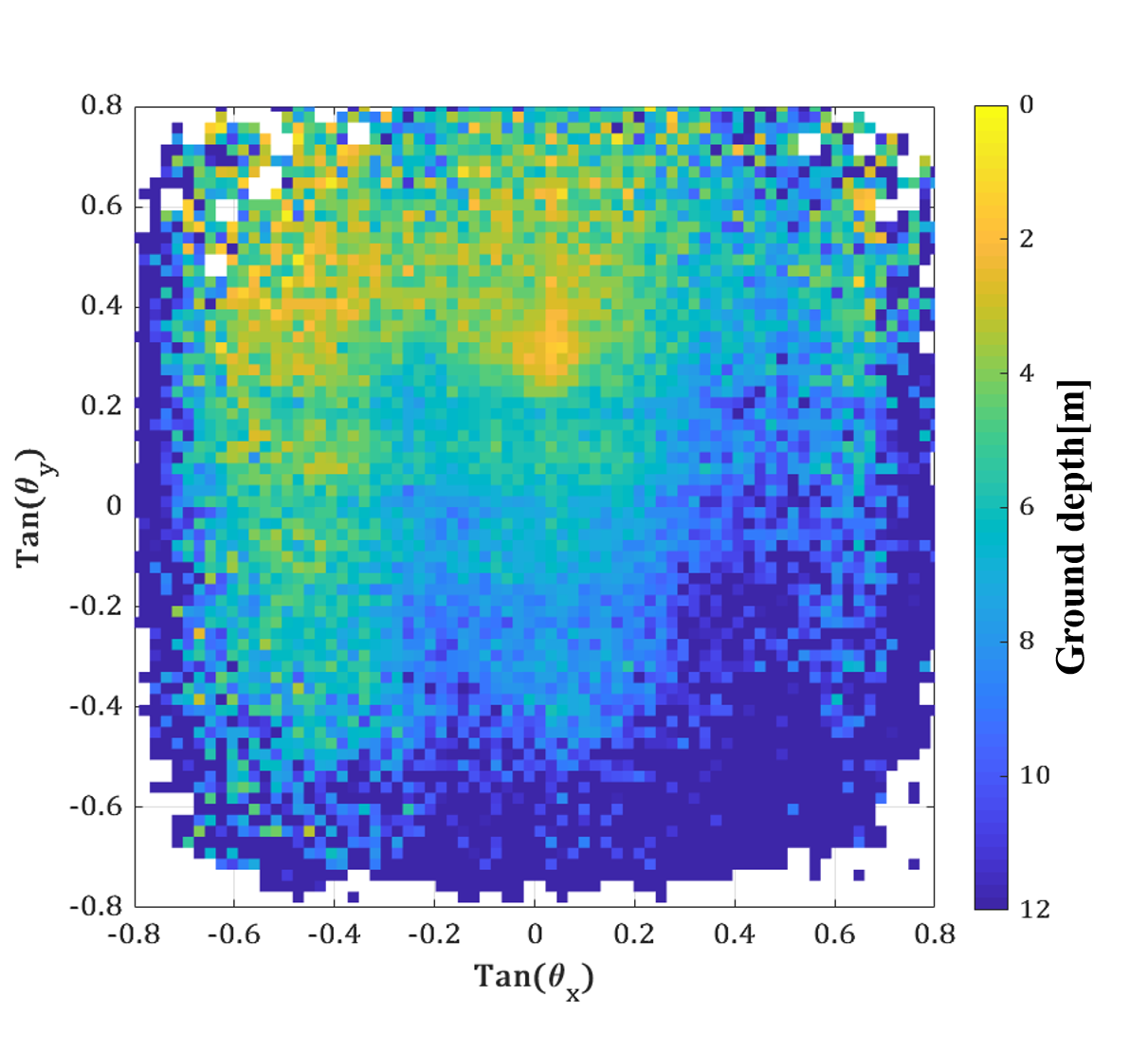}
  \caption{The angular ground depth \(D(\tau_x, \tau_y)\) derived by applying the lookup tables from muon transport simulations to the measured data.}
  \label{fig:angular_depth}
\end{figure}

By applying the lookup tables from muon transport simulations to the measured data, we derive the angular ground depth  \(D(\tau_x, \tau_y)\), which acts as a proxy for the integrated density of the overburden. Fig.~\ref{fig:angular_depth} shows the resulting density map. This map, the primary result of our study, reveals variations in the overburden that may indicate subsurface anomalies, such as voids or high-density inclusions.

-----
\section{Discussion}

The measurement shows that muons can map underground features and resolve structural variations in the overburden. The angular muon flux distribution (Fig.~\ref{fig:2}) highlights anisotropies consistent with the ventilation shafts seen in the LiDAR scan (Fig.~\ref{fig:pit}), which was integrated into our simulation framework. The match between simulated and measured angular patterns confirms that the features correspond to actual geometric structures above the detector. However, while the relative angular trends are meaningful, the overall flux normalization is affected by several factors: detector efficiency, the assumption of uniform overburden density, and the flat surface profile approximation above the cistern, all of which differ from actual site conditions. These discrepancies are discussed below.

First, the detector efficiency is low (less than 20\%), mainly because of a suboptimal trigger scheme and excessively high detection thresholds. This inefficiency leads to a significant under-counting of muon events and a reduced flux compared to simulation predictions. Improving the trigger logic and recalibrating the thresholds are essential steps to enhance the measurement’s sensitivity and accuracy.

Second, the simulation of the atmospheric muon flux assumed a sea-level model. Since Jerusalem is approximately 650~m above sea level, the actual muon flux is higher, with a slightly different angular distribution. Empirical models of cosmic-ray muon production~\cite{Reyna2006ASP,Bugaev1998} indicate that at this elevation, the vertical muon flux increases by approximately 15–20\% compared to sea level. This increase results from the thinner atmosphere, which allows more muons to reach the surface. Not applying altitude correction causes a systematic difference in both the absolute muon rate and its angular distribution, thus impacting the inferred overburden.

Third, although the cistern's interior was accurately scanned with LiDAR, the simulations assumed a flat ground surface above it. In reality, the terrain has significant features on the meter scale, such as large stone walls, mature olive trees, and nearby buildings. These were not included in the simulation geometry, and their omission adds extra error in modeling muon attenuation.

Finally, limitations in muon transport simulations also contribute to the observed discrepancies. These simulations, based on the energy loss equation (Eq.~\ref{eq.loss}), incorporate ionization and radiative losses via energy-dependent coefficients \(a(E)\) and \(b(E)\). The lookup tables derived from these models assume a homogeneous material and simplified geometry, neglecting local inhomogeneities or complex surface structures. This simplification can cause local errors in the reconstructed density map. Additional uncertainties come from environmental variations and potential systematic biases in detector calibration. Ongoing efforts include adding altitude corrections, improving calibration procedures, and expanding the detector array to enable full three-dimensional subsurface imaging.

Furthermore, as part of our forward-looking strategy, we plan to incorporate AI-based image processing techniques to reduce noise, improve flux map reconstruction, and detect subtle anomalies in challenging data conditions. These efforts will build on advances in other fields such as medical tomography and geophysical imaging, and we believe they have strong potential for archaeological muon imaging as well.

\section{Conclusion}

This study demonstrates the utility of muon imaging as an effective, non-invasive technique for mapping underground features, specifically in the archaeological context of the cistern in the City of David, Jerusalem. We could derive key insights into the subsurface density variation and overburden characteristics by deploying a custom muon detector and integrating high-resolution LiDAR mapping. While discrepancies between the measured and expected muon flux were identified, primarily attributed to detector efficiency, altitude effects, and the inherent complexity of the overlying structures, these findings pave the way for further refinement and optimization of the muon detection process. Future work will address these uncertainties by improving calibration methods, incorporating altitude adjustments in simulations, and expanding the detector network to enable comprehensive three-dimensional imaging. Overall, this study highlights the potential of muon tomography for archaeological investigations and supports its continued application in exploring buried structures and understanding ancient civilizations.
\section{Data Availability}
The data supporting the findings of this study are available from the corresponding author upon reasonable request.
Due to the technical nature of the data and its dependence on specific analysis procedures, effective use requires further explanation from the authors.
\begin{acknowledgments}

This work was supported in part by the Pazy and Cogito foundations.

The site is located in the City of David within the Jerusalem Walls National Park. Y. Shalev and F. Vukosavović excavated on behalf of the IAA, Permit No. 8928.
\end{acknowledgments}
\vspace{2 cm}

\section*{References}
\bibliography{main.bib}

\end{document}